\documentclass[]{emulateapj}
\usepackage{amsmath}
\usepackage{graphicx}
\usepackage{multirow}
\usepackage{color}
\usepackage{latexsym}
\usepackage{amssymb}
\usepackage{epsfig}
\usepackage{verbatim}
\usepackage{ifpdf}

\begin{document}

\title{Hot-Jupiter Inflation due to Deep Energy Deposition}

\author{Sivan Ginzburg and Re'em Sari}

\affil{Racah Institute of Physics, The Hebrew University, Jerusalem 91904, Israel}

\begin{abstract}
Some extrasolar giant planets in close orbits---``hot Jupiters''---exhibit larger radii than that of a passively cooling planet.
The extreme irradiation $L_{\rm eq}$ these hot Jupiters receive from their close in stars creates a thick isothermal layer in their envelopes, which slows down their convective cooling, allowing them to retain their inflated size for longer. This is yet insufficient to explain the observed sizes of the most inflated planets. 
Some models invoke an additional power source, deposited deep in the planet's envelope. Here we present an analytical model for the cooling of such irradiated, and internally heated gas giants. We show that a power source $L_{\rm dep}$, deposited at an optical depth $\tau_{\rm dep}$, creates an exterior convective region, between optical depths $L_{\rm eq}/L_{\rm dep}$ and $\tau_{\rm dep}$, beyond which a thicker isothermal layer exists, which in extreme cases may extend to the center of the planet. This convective layer, which occurs only for $L_{\rm dep}\tau_{\rm dep}>L_{\rm eq}$, further delays the cooling of the planet. Such a planet is equivalent to a planet irradiated with $L_{\rm eq}\left(1+L_{\rm dep}\tau_{\rm dep}/L_{\rm eq}\right)^\beta$, where $\beta\approx 0.35$ is an effective power-law index describing the radiative energy density as function of the optical depth for a convective planet $U\propto\tau^\beta$. Our simple analytical model reproduces the main trends found in previous numerical works, and provides an intuitive understanding. We derive scaling laws for the cooling rate of the planet, its central temperature, and radius. These scaling laws can be used to estimate the effects of tidal or Ohmic dissipation, wind shocks, or any other mechanism involving energy deposition, on sizes of hot Jupiters.
\end{abstract}


\section{Introduction}
\label{sec:introduction}

Hot Jupiters are a class of giant planets that orbit their parent stars much closer than Jupiter orbits the sun, therefore exhibiting high surface temperatures \citep[see, e.g,][]{MayorQueloz95}. The observed radii of many of these exoplanets are greater than Jupiter's radius $R_J$, with the largest planets reaching radii $\sim 2R_J$ \citep{Baraffe2010,Anderson2011,Chan2011,Hartman2011,SpiegelBurrows2013}. This discovery challenges theoretical evolution models, which predict that gas giants cool and contract to smaller radii, closer to $1.0R_J$, at their inferred age, which is often greater than 1Gyr \citep{Burrows2000,Burrows2007,Laughlin2005}. 
To some extent, the larger radii are the result of the intense stellar irradiation, which induces a deep radiative, nearly isothermal, layer at the outer edge of the otherwise fully convective planets \citep{Guillot96,ArrasBildsten2006}. This radiative layer slows down the evolutionary cooling of the planet significantly, in comparison with isolated planets \citep{Burrows2000,Chabrier2004}. The slower cooling results in a higher bulk entropy at a given age, and therefore, a larger radius \citep[see, e.g,][]{ArrasBildsten2006,SpiegelBurrows2012,MarleauCumming2014}.

Yet, the radii of at least some of the inflated hot Jupiters exceed theoretical evolutionary models, even with stellar irradiation taken into account \citep{Baraffe2003,Burrows2007,Liu2008}. To settle this discrepancy, a number of explanations have been suggested \citep[see][for comprehensive reviews]{Baraffe2010,FortneyNettelmann2010,SpiegelBurrows2013,Baraffe2014}. The different explanations may be divided into several categories. While some works study the effects of extra power, deposited at some depth inside the planet, other focus on enhanced atmospheric opacities \citep{Burrows2007}, suppression of convective
heat loss \citep[double-diffusive convection, see][]{ChabrierBaraffe2007,LeconteChabrier2012}, turbulent mixing \citep{YoudinMitchell2010}, or more consistent coupling of the day side and night side of the planet, and taking into account the three-dimensional nature of the problem  \citep{Hansen2008,Guillot2010,SpiegelBurrows2010,SpiegelBurrows2013,Budaj2012,Dobbs-Dixon2012,Perez-BeckerShowman2013,Mayne2014,RauscherShowman2014}.

In this work we focus on the effects of extra power sources deep in the planet. Possible heat sources include tidal dissipation due to orbital eccentricity \citep{Bodenheimer2001,Bodenheimer2003,Gu2003,WinnHolman2005,Jackson2008,Liu2008,IbguiBurrows2009,Miller2009,Ibgui2010,Ibgui2011,Leconte2010}, ``thermal tides'' \citep{ArrasSocrates2009a,ArrasSocrates2009b,ArrasSocrates2010,Socrates2013}, Ohmic heating \citep{BatyginStevenson2010,Perna2010,Perna2012,Batygin2011,HuangCumming2012,RauscherMenou2013,WuLithwick2013,RogersShowman2014}, and conversion of a portion of the absorbed stellar flux into kinetic energy in the atmosphere, which is then dissipated at a greater depth \citep{GuillotShowman2002,ShowmanGuillot2002}. While some of these mechanisms (e.g. Ohmic heating, kinetic energy dissipation) are driven by the stellar irradiation, others (e.g. tidal dissipation) are not. If, as some studies suggest, excess planet inflation is correlated with the stellar irradiation \citep{Laughlin2011,Schneider2011}, then the former class of heat deposition mechanisms is favored.

We do not attempt to determine which power sources are more plausible. Alternatively, in the current work we study the effects of a general power source on the cooling, and therefore the radius, of a hot Jupiter. Specifically, we are interested in the dependence of the planet's cooling history on the amount of additional power, and on the depth at which this power is deposited inside the planet. Recently, \citet{SpiegelBurrows2013} have conducted such a study numerically \citep[see also][]{GuillotShowman2002,Baraffe2003}. In this work we present a simple analytical model for the cooling of insolated gas giants in the presence of heat deposition somewhere in their envelope. Despite our simplifications, this model describes the essential processes, and provides an intuitive explanation to the main trends of the scenario.

The outline of the paper is as follows. In Section \ref{sec:irradiated} we present our model for irradiated planets, without additional heat sources. The effects of additional deposited heat are studied in Sections \ref{sec:deposition} and \ref{sec:radius}. In Section \ref{sec:comparison} we compare our results to the calculations of \citet{SpiegelBurrows2013}, and the main conclusions are summarized in Section \ref{sec:conclusions}.

\section{Irradiated Planets}
\label{sec:irradiated}

In this section we present our simplified ``toy model'' for irradiated gas giants, without additional heat sources. The model is one dimensional, and does not differentiate between the day and night sides of the planet \citep[see][]{SpiegelBurrows2013}.
Many aspects of the deposition-free model are essentially similar to a previous analytical work by \cite{ArrasBildsten2006}, and are also implemented by \citet{YoudinMitchell2010}.

Isolated gas giants, such as Jupiter, are almost fully convective, with convection beginning at an optical depth $\tau\sim 1$ \citep{Trafton64,Trafton67,Hubbard68,StevensonSalpeter77}; see also \citet{Guillot94} for a reexamination. Insolated planets, on the other hand, are expected to develop a deep radiative envelope \citep{Guillot96}.

\subsection{Convective Interior Structure}
\label{sec:convective}

The analysis of irradiated planets is simpler when considering their profiles in the $(\tau,U)$ plane, with
\begin{equation}\label{eq:tau}
\tau(r)=\int_r^R{\kappa\rho dr'}
\end{equation}
denoting the optical depth at radius $r$, from the planet surface, at radius $R$, and $U\equiv aT^4$ is the radiative energy density. The density, temperature, and opacity are denoted by $\rho$, $T$, and $\kappa$, respectively, and $a$ is the radiation constant. 
Our main simplification is to assume that this profile can be approximated by a power law in the convective interior
\begin{equation}\label{eq:U_con}
\frac{U}{U_c}=\left(\frac{\tau}{\tau_c}\right)^\beta,
\end{equation} 
 with $U_c\equiv a T_c^4 $ denoting the central radiation energy density, determined by the central temperature $T_c$.

This scaling can be derived in the case of a power-law opacity $\kappa\propto\rho^a T^b$, as assumed by \citet{ArrasBildsten2006} and \citet{YoudinMitchell2010}. The opacity is combined with the adiabatic temperature and pressure relation in the convective region \citep[see, e.g.,][]{Hubbard77,Saumon96} $T\propto P^{1/(n+1)}$, and the resulting polytropic density profile at the edge of the planet $\rho\approx\rho_c(1-r/R)^n$ \citep[$\rho_c$ denotes the central density; see, e.g.,][]{MatznerMcKee99}. We neglect super-adiabatic corrections to the convective temperature profile \citep{Trafton67,ChabrierBaraffe2007}.
These power laws yield
\begin{equation}\label{eq:beta}
\beta=\frac{4}{b+1+n(a+1)},
\end{equation}
and $\tau_c\sim\kappa_c\rho_c R$, with $\kappa_c$ denoting the estimate for the central opacity, if the power-law opacity could be continued to the center.

The Rosseland opacity (we assume a gray approximation) in the range of interest is depicted in Figure 1 of \cite{ArrasBildsten2006}. The molecular dominated opacity below $T_0\approx 2500\rm K$ is fit by $a=b=0.5$. Above $T_0$, the opacity rises steeply, with $b=7$ ($a=0.5$), due to $\rm H^-$, whose population increases primarily as a consequence of metal ionization \citep[see, e.g.,][]{KippenhahnWeigert94}. At even higher temperatures, above $\sim 10^4\rm K$, the $\rm H^-$ vanishes, due to lack of neutral hydrogen, and the opacity is given by the Kramers bound-free opacity law, with $a=1$, $b=-3.5$ \citep[see, e.g.,][]{KippenhahnWeigert94, Padmanabhan2000}. 

As explained in Sections \ref{sec:transition} and \ref{sec:luminosity}, the internal luminosity of a planet is determined by the location of the radiative-convective boundary. The optical depth (and pressure level) of this boundary is given by the opacity power-law which is suitable for the (almost isothermal) radiative-zone temperature, and does not depend on the deeper opacity structure. For planets without additional heating, this temperature is typically below $T_0$, and we therefore get $\beta=0.8$, by taking $n\approx 2$ \citep{Peebles64,Hubbard77,Saumon96}. Throughout the majority of this work we adopt the approximation of small inflations relative to the zero-temperature radius (see Section \ref{sec:radius}), implying roughly constant central densities and pressures. Consequently, a constant $\tau_c=\tau_0\sim 10^{11}$ is suitable for a Jupiter-size planet. Inspection of numerical evolution models \citep[see Table 1 in][]{Burrows97} reveals that the ratio between central and photosphere temperatures, which is given by setting $\tau\sim 1$ in Equation \eqref{eq:U_con}, $T_c/T_{\rm ph}=\tau_c^{\beta/4}$, is $\sim 100$, and changes only by a factor of 2 during the cooling of a Jupiter mass isolated planet, consistent with our estimate. 

In Section \ref{sec:deposition} we show that heat deposition raises the radiative-convective boundary temperature above $T_0$, requiring a more general form, which takes into account the strong dependence of the $\rm H^-$ opacity on the temperature
\begin{equation}\label{tau_c}
\frac{\tau_c}{\tau_0}=\left(\frac{T_c}{T_0}\right)^b.
\end{equation}    
In this case we get $\beta=0.35$ from Equation \eqref{eq:beta}, which is suitable for temperatures above $T_0$. As we show in Section \ref{sec:transition}, the convective profile must satisfy $\beta<1$.

\clearpage

\subsection{Radiative Envelope Structure}
\label{sec:radiative}

Proximity to a star dictates an equilibrium temperature $T_{\rm eq}$ on the planet surface, which we take as the photosphere, at $\tau\sim 1$. Our model is spherical so we assume an even distribution of the absorbed heat over the planet surface, for which \cite[see, e.g.,][]{Guillot96} 
\begin{equation}\label{eq:Teq}
T_{\rm eq}=(1-A)^{1/4} T_\sun\left(\frac{R_\sun}{2D}\right)^{1/2},
\end{equation}
with $T_\sun$ and $R_\sun$ denoting the stellar temperature and radius, respectively, and $D$, $A$ are the planet's orbital distance and albedo, respectively. This equilibrium temperature defines an energy density $U_{\rm eq}\equiv aT_{\rm eq}^4$, and luminosity $L_{\rm eq}\equiv 4\pi R^2\sigma T_{\rm eq}^4$, where $\sigma$ is the Stefan-Boltzmann constant.

The radiation energy is related to the internal luminosity of the planet $L_{\rm int}$ through the diffusion approximation
\begin{equation}\label{eq:diff}
\frac{L_{\rm int}}{4\pi R^2}=
\frac{c}{3}\frac{dU}{d\tau},
\end{equation}
where $c$ is the speed of light. Close to the surface, $L_{\rm int}$ is constant (in space), since it is the result of cooling of the entire planet, and the heat capacity of the outer layers is negligible. Therefore, we can integrate Equation \eqref{eq:diff} and find the radiative profile
\begin{equation}\label{eq:U_rad}
U=U_{\rm eq}+\frac{3}{c}\frac{L_{\rm int}}{4\pi R^2}\tau,
\end{equation}  
where the distinction between the boundary conditions $U=U_{\rm eq}$ at $\tau=0$ and $\tau\sim 1$ is negligible, since we are generally interested in $\tau\gg 1$. \citet{Bodenheimer2003} use a similar boundary condition (with $U=U_{\rm eq}$ at $\tau=2/3$), while \citet{ArrasBildsten2006} and \citet{YoudinMitchell2010} define a temperature $T_{\rm deep}$ for their boundary condition, which is correlated with $T_{\rm eq}$, therefore leading to similar conclusions. More elaborate boundary conditions, which incorporate a frequency-dependent opacity, and require the solution of the radiative transfer equation, or the use of a two-stream approximation \citep{Barman2001,Hubeny2003,Hansen2008,Guillot2010,GuillotHavel2011}, are beyond the scope of this work, and their effect on the temperature deep inside the radiative zone (and therefore on the cooling rate) can be modeled by a somewhat different effective $T_{\rm eq}$. We remark that an atmosphere which is more transparent in the optical regime \cite[although this does not seem to be the case for hot Jupiters; see, e.g.,][]{SpiegelBurrows2013} may induce a greenhouse effect, which can be treated with our model as heat deposited at $\tau>1$, where the depth for optical light is unity.

Planar symmetry, and the focus on the outer layers of the planet throughout this work, are justified both self-consistently in Section \ref{sec:luminosity}, and by \citet{Guillot96}, who find that during the 10Gyr evolution of a hot Jupiter, the radiative region penetrates only a few percent of the radius.

\subsection{Radiative-Convective Transition}
\label{sec:transition}

According to the Schwarzschild criterion, when the radiative temperature gradient becomes larger than the adiabatic gradient, convective instability develops. 
Differentiating Equations \eqref{eq:U_con} and \eqref{eq:U_rad} shows that convection sets in when
\begin{equation}\label{eq:transition}
\frac{4}{3}\frac{L_{\rm eq}}{L_{\rm int}}\frac{1}{\tau}<\frac{1-\beta}{\beta}.
\end{equation} 
We see from Equation \eqref{eq:transition} that at $\tau\to 0$ the profile is radiative. We also see that to ensure convection at $\tau\to\infty$, we must have $\beta<1$.

\ifpdf

\else

\begin{figure}[tbh]
\epsscale{1} \plotone{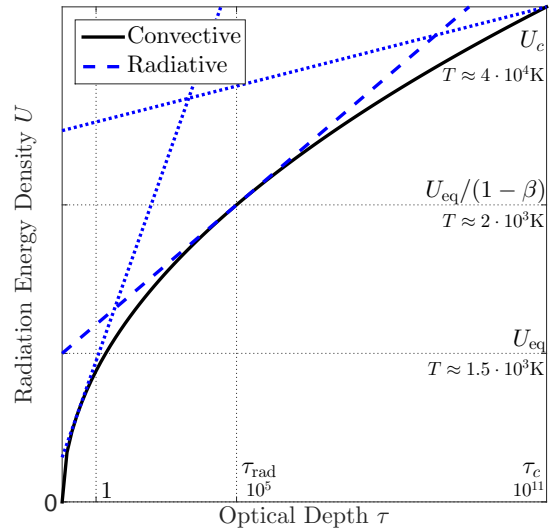}
\caption{Transition between convective (solid black line) and radiative (dashed blue line) profiles. The convective profile follows Equation \eqref{eq:U_con}, while the radiative profile is according to Equation \eqref{eq:U_rad}. The combined profile follows the radiative profile for $\tau<\tau_{\rm rad}$, and the convective profile for $\tau>\tau_{\rm rad}$. Two additional extreme cases are plotted for the same values of $U_c$ and $\tau_c$ (dotted blue lines). A very high value $U_{\rm eq}>U_c(1-\beta)$ corresponds to a fully radiative planet, while a very low value of $U_{\rm eq}$ leads to $\tau_{\rm rad}<1$, and therefore a fully convective planet. Typical values of the optical depth and temperature are given.
\label{fig:transition}}
\end{figure}

\fi

The transition between convective and radiative profiles is illustrated schematically in Figure \ref{fig:transition}. For a planet with given interior ($U_c$, $\tau_c$) and exterior ($U_{\rm eq}$) boundary conditions, the convective profile is given by Equation \eqref{eq:U_con}, while the radiative profile is given by Equation \eqref{eq:U_rad}. Since the radiative profile is linear in $\tau$, the luminosity $L_{\rm int}$ and the radiative-convective transition point $\tau_{\rm rad}$ are calculated by finding the tangent to the adiabatic profile. For $\tau<\tau_{\rm rad}$, the radiative gradient is smaller than the adiabatic, so the profile is radiative. For $\tau>\tau_{\rm rad}$, the radiative gradient is larger than the adiabatic, so the profile is convective. Quantitatively, the tangency point is found by equating Equations \eqref{eq:U_con} and \eqref{eq:U_rad} and their derivatives. The solution for $\tau_{\rm rad}$ is
\begin{equation}\label{eq:tau_rad}
\left(\frac{\tau_{\rm rad}}{\tau_c}\right)^\beta=\frac{1}{1-\beta}\frac{U_{\rm eq}}{U_c},
\end{equation}
which corresponds to a radiation energy density
\begin{equation}\label{eq:u_tau_rad}
U(\tau_{\rm rad})=\frac{1}{1-\beta}U_{\rm eq},
\end{equation}
implying that the radiative zone is isothermal to within a factor of $(1-\beta)^{-1/4}\approx 1.5$. Equation \eqref{eq:tau_rad} indicates that the radiative region thickens with increasing solar irradiation and with decreasing central temperature \citep[see also][]{Guillot96,Burrows2000,ArrasBildsten2006,YoudinMitchell2010}. 

\subsection{Internal Luminosity and Cooling}
\label{sec:luminosity}

Equations \eqref{eq:transition} and \eqref{eq:tau_rad} connect the internal luminosity to the central temperature
\begin{equation}\label{eq:L_U}
\frac{L_{\rm int}}{L_{\rm eq}}\sim\frac{1}{\tau_{\rm rad}}\sim\frac{1}{\tau_c}\left(\frac{U_c}{U_{\rm eq}}\right)^{1/\beta}.
\end{equation}

Note that Equation \eqref{eq:L_U} breaks down when $\tau_{\rm rad}\lesssim 1$.  In this case, the planet is considered to be fully convective, and its luminosity is determined by the radiation energy density at the photosphere $U_{\rm ph}$. Substituting $\tau=1$ in Equation \eqref{eq:U_con} we find
\begin{equation}\label{eq:Uph}
U_{\rm ph}=U_c\left(\frac{1}{\tau_c}\right)^\beta,
\end{equation}
and luminosity 
\begin{equation}\label{eq:L_fully_conv}
\frac{L_{\rm int}}{L_{\rm eq}}=\frac{U_{\rm ph}}{U_{\rm eq}}=\left(\frac{1}{\tau_c}\right)^\beta\frac{U_c}{U_{\rm eq}}.
\end{equation}
Comparing Equations \eqref{eq:tau_rad} and \eqref{eq:Uph} shows that the condition $\tau_{\rm rad}\lesssim 1$ is equivalent to the condition $U_{\rm ph}\gtrsim U_{\rm eq}$. In other words, as long as the central temperature of the planet is high enough to ensure that a convective profile will reach $\tau\sim 1$ without dropping below $T_{\rm eq}$, the planet remains fully convective. However, as the planet cools, the central temperature decreases, and the convective profile would have predicted temperatures below $T_{\rm eq}$. This situation is unphysical, and an approximately isothermal radiative zone develops.

To summarize, using Equations \eqref{eq:L_U} and \eqref{eq:L_fully_conv}, the internal luminosity of the planet, as a function of its central radiation energy density is given by
\begin{equation}\label{eq:Lint_cases}
\frac{L_{\rm int}}{L_{\rm eq}}\sim
\begin{cases}
\frac{1}{\tau_c}\left(\frac{U_c}{U_{\rm eq}}\right)^{1/\beta} & \frac{U_c}{U_{\rm eq}}<\tau_c^\beta
\\[1.5ex]
\left(\frac{1}{\tau_c}\right)^\beta\frac{U_c}{U_{\rm eq}} &
\frac{U_c}{U_{\rm eq}}>\tau_c^\beta
\end{cases},
\end{equation}
where the transition is when $\tau_{\rm rad}\sim L_{\rm eq}/L_{\rm int}\sim 1$. Since $\beta<1$, Equation \eqref{eq:Lint_cases} indicates that the proximity to a star, which is manifested for $U_c/U_{\rm eq}<\tau_c^\beta$, diminishes the internal luminosity and slows down the planet's cooling \citep[see also][]{Guillot96,Burrows2000,ArrasBildsten2006,YoudinMitchell2010,SpiegelBurrows2013}.

As a planet cools, it makes the transition from an isolated planet, which is not affected by stellar irradiation, and matches the high $U_c$ limit of Equation \eqref{eq:Lint_cases}, to an insolated planet, which matches the low $U_c$ limit. The age, at which this transition takes place, may be estimated by writing an evolution equation for the central temperature of an isolated planet
\begin{equation}\label{eq:cooling}
L_{\rm int}\sim-k_B\frac{M}{m_p}\frac{dT_c}{dt},
\end{equation}
where $k_B$ is Boltzmann's constant, $m_p$ is the proton mass, $M$ is the mass of the planet, and $t$ denotes time. Equation \eqref{eq:cooling} is valid for the cooling degenerate phase of the planet's evolution, which follows a rapid non-degenerate contraction phase \citep[see][]{Guillot2005,ArrasBildsten2006}. Substituting the luminosity from Equation \eqref{eq:Lint_cases}, we get
\begin{equation}\label{eq:cooling_T}
\frac{4\pi R^2\sigma T_c^4}{\tau_c^\beta}\sim-k_B\frac{M}{m_p}\frac{dT_c}{dt}.
\end{equation} 
Assuming that the planet cools down from high temperatures, Equation \eqref{eq:cooling_T} predicts the cooling time to a temperature $T_c=T_{\rm eq}\tau_c^{\beta/4}$, which marks the transition to the insolated regime (at this stage the photosphere reaches a temperature $T_{\rm ph}\sim T_{\rm eq}$):
\begin{equation}\label{eq:cooling_time}
t\sim\frac{M}{m_p}\frac{k_B}{12\pi R^2\sigma T_{\rm eq}^3}\tau_c^{\beta/4}\approx 7\textrm{Myr}\left(\frac{10^3\rm K}{T_{\rm eq}}\right)^3,
\end{equation}
were the numerical value is estimated for $M_J$ (Jupiter mass) and $R_J$. This estimate roughly fits Jupiter, which is a few Gyr old \citep{Guillot2005}, has an equilibrium temperature $T_{\rm eq}\approx 110\rm K$, and is close to the transition, since $T_{\rm ph}\approx 120\rm K$ \citep{Hanel81}. Strongly irradiated planets orbit their parent star closer than Jupiter by a factor of 100, thus having, using Equation \eqref{eq:Teq}, $T_{\rm eq}\sim 10^3\rm K$, and therefore reach the insolated regime and develop a radiative zone after less than $10^7$ years \citep[consistent with][]{Guillot96}.

In the following, insolated stage of planetary evolution, Equation \eqref{eq:cooling} is combined with the low $U_c$ limit of Equation \eqref{eq:Lint_cases} to produce a cooling equation
\begin{equation}\label{eq:cooling_insolated}
\begin{split}
T_c(t)\sim\left[\tau_ck_B\frac{M}{m_p}\frac{T_{\rm eq}^{4(1-\beta)/\beta}}{4\pi R^2\sigma}\frac{1}{t}\right]^{\beta/(4-\beta)}\\\approx 
6\cdot 10^4{\rm K}\left(\frac{T_{\rm eq}}{10^3\rm K}\right)^{1/4}\left(\frac{1{\rm Gyr}}{t}\right)^{1/4}
,
\end{split}
\end{equation}
with the numerical values estimated for $M_J$, $R_J$. This result is consistent with \citet{Guillot96}. At an age of 8Gyr, the radiative region encompasses a few percent in radius, $1-r/R\sim(\rho/\rho_c)^{1/n}\sim T_{\rm eq}/T_c$, consistent with \citet{Guillot96}, and self-consistently justifying our focus on the outer layers of the planet.

A third stage in a cooling planet's life, equilibrium, is reached when $U_c=U_{\rm eq}$. In this case, according to Equation \eqref{eq:tau_rad}, the planet becomes fully radiative, and the luminosity vanishes $L_{\rm int}=0$ (since the temperature gradient vanishes). This condition determines the planet's equilibrium central temperature, and therefore its equilibrium radius. More precisely, the planet becomes fully radiative when $U_c$ drops below $U_{\rm eq}/(1-\beta)$. It is easy to see, from Equation \eqref{eq:cooling_insolated}, that without additional heat deposition, planets never reach equilibrium (the timescales are too long).

The three stages of planetary evolution, described in this section, are displayed in Figure \ref{fig:transition}, for planets with the same central temperature and optical depth, but different equilibrium temperatures. 

\section{Deposition of Additional Power}
\label{sec:deposition}

In this section we discuss how the internal luminosity is affected by an additional heat source, which deposits power at some depth inside the planet. Our goal is to find $L_{\rm int}(T_{\rm eq},T_c,L_{\rm dep},\tau_{\rm dep})$, with $L_{\rm dep}$ denoting the deposited luminosity, and $\tau_{\rm dep}$ denoting the optical depth where the power is deposited. In Section \ref{sec:luminosity}, and specifically in Equation \eqref{eq:Lint_cases}, we found $L_{\rm int}(L_{\rm dep}=0)$, which we abbreviate hereafter as $L_{\rm int}^0$. We focus on planets which have developed a substantial radiative zone, $\tau_{\rm rad}>1$, so $L_{\rm int}^0$ is determined by the low $U_c$ limit of Equation \eqref{eq:Lint_cases}. The motivation is that strongly irradiated planets develop such a radiative zone in less than $10^7$ years, as explained in Section \ref{sec:luminosity} \citep[see also][]{Guillot96}.

Without heat sources, the radiative zone is described by Equation \eqref{eq:U_rad}. The addition of a deposited power $L_{\rm dep}$ at $\tau_{\rm dep}$ introduces a jump in the profile slope. Explicitly, the radiative profile is altered:
\begin{equation}\label{eq:deposit_rad}
\frac{dU}{d\tau}=\frac{3}{4\pi R^2c}\cdot
\begin{cases}
L_{\rm tot}\equiv L_{\rm int}+L_{\rm dep} & \tau<\tau_{\rm dep}
\\[1.5ex]
L_{\rm int} & \tau>\tau_{\rm dep}
\end{cases},
\end{equation}
with $L_{\rm tot}$ denoting the total luminosity for $\tau<\tau_{\rm dep}$.
The radiative profile, with deposited heat included, is displayed in Figure \ref{fig:deposit}. As seen in Figure \ref{fig:deposit}, the deposited heat effectively changes $U_{\rm eq}$ to some higher $U_{\rm eq}^{\rm eff}$. The internal luminosity $L_{\rm int}$ is found, as in Section \ref{sec:transition}, by calculating the tangent from the point $[\tau_{\rm dep},U(\tau_{\rm dep})]$ to the convective profile, which is equivalent to calculating the tangent from the point $[0,U_{\rm eq}^{\rm eff}]$. Therefore, the results of Section \ref{sec:transition} are reproduced, but with $U_{\rm eq}^{\rm eff}$ instead of $U_{\rm eq}$. Since, from Equation \eqref{eq:L_U}, $L_{\rm int}\propto U_{\rm eq}^{-(1-\beta)/\beta}$, the internal luminosity decreases. This result is also evident graphically in Figure \ref{fig:deposit}: the slope of the radiative tangent decreases, as well as the deeper penetration of the radiative zone into the planet interior.

\subsection{Secondary Convective Region}
\label{sec:secondary}

In Figure \ref{fig:deposit} we assumed that the outer layers remain radiative. However, if the total power, $L_{\rm tot}$, which has to be evicted from the planet, is high enough, then it cannot be delivered radiatively, and a secondary convective instability develops. According to Equations \eqref{eq:transition} and \eqref{eq:deposit_rad}, with $L_{\rm tot}$ replacing $L_{\rm int}$ as the evicted power, convection appears at an optical depth
\begin{equation}\label{eq:tau_b}
\tau_b=\frac{4}{3}\frac{\beta}{1-\beta}\frac{L_{\rm eq}}{L_{\rm tot}},
\end{equation}
where the radiation energy density reaches $U_{\rm eq}/(1-\beta)$.
If we focus on intense deposition $L_{\rm dep}\gg L_{\rm int}$, then we may approximate $L_{\rm tot}\approx L_{\rm dep}$, and using Equation \eqref{eq:tau_b}, $\tau_b\sim L_{\rm eq}/L_{\rm dep}$.

\ifpdf

\else

\begin{figure}[tbh]
\epsscale{1} \plotone{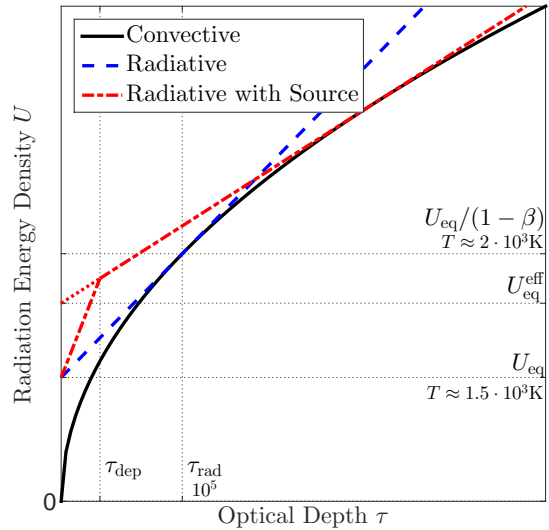}
\caption{Modification of the radiative profile due to the deposition of heat at optical depth $\tau_{\rm dep}$. The convective profile (solid black line), and the radiative profile without a heat source (dashed blue line) are similar to Figure \ref{fig:transition}. The modified radiative profile (dot-dashed red line) is described by Equation \eqref{eq:deposit_rad}. The imaginary continuation of the profile with a slope $3L_{\rm int}/(4\pi R^2 c)$ down to $\tau=0$ (dotted red line) defines $U_{\rm eq}^{\rm eff}$. As in Figure \ref{fig:transition}, the combined profile follows the radiative profile up to the tangency point, and then it follows the convective profile.
\label{fig:deposit}}
\end{figure}

\begin{figure}[tbh]
\epsscale{1} \plotone{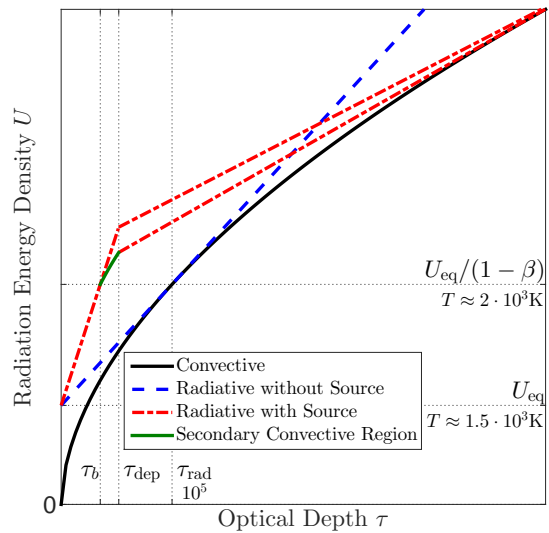}
\caption{Emergence of a secondary convective region from $\tau_b$ to $\tau_{\rm dep}$ (solid green line). The convective (solid black line) and radiative (dashed blue line) profiles without a heat source are similar to Figures \ref{fig:transition} and \ref{fig:deposit}. The radiative profile with heat deposited at $\tau_{\rm dep}$ (upper dot-dashed red line) is similar to Figure \ref{fig:deposit}, and is inadequate in this case. Due to the secondary convective region, the radiative profile is altered, and it follows the lower dot-dashed red line, up to the tangency point with the main convective profile.
\label{fig:secondary}}
\end{figure}

\fi

The temperature profile, with deposited heat added, can be divided into two cases, depending on $L_{\rm dep}$ and $\tau_{\rm dep}$. If $\tau_{\rm dep}<\tau_b$, or equivalently $L_{\rm dep}\tau_{\rm dep}/L_{\rm eq}\lesssim 1$, The depiction in Figure \ref{fig:deposit} is accurate. As seen in Figure \ref{fig:deposit}, in this case $U_{\rm eq}^{\rm eff}$ is restricted by a maximal value $U_{\rm eq}^{\rm eff}<U_{\rm eq}/(1-\beta)$, limiting the reduction in internal luminosity, using Equation \eqref{eq:L_U}, to an insignificant effect $L_{\rm int}/L_{\rm int}^0>(1-\beta)^{(1-\beta)/\beta}\sim 1$.

In the case where $\tau_{\rm dep}>\tau_b$, a secondary convective region emerges between $\tau_b$ and $\tau_{\rm dep}$. This region is displayed in Figure \ref{fig:secondary}. The modified luminosity can be calculated by finding $U_{\rm eq}^{\rm eff}$. Using Equation \eqref{eq:tau_b}, approximating $L_{\rm dep}\gg L_{\rm int}$, and noting that in this regime $L_{\rm dep}\tau_{\rm dep}/L_{\rm eq}\gtrsim 1$, we find 
\begin{equation}\label{eq:U_eq_eff_big}
\frac{U_{\rm eq}^{\rm eff}}{U_{\rm eq}}\approx\frac{1}{1-\beta}\frac{U(\tau_{\rm dep})}{U(\tau_b)}\sim\left(\frac{L_{\rm dep}\tau_{\rm dep}}{L_{\rm eq}}\right)^\beta,
\end{equation}
where $U(\tau_{\rm dep})$ is adiabatically related to $U(\tau_b)$.
Combining this result with Equation \eqref{eq:L_U}, we get $L_{\rm int}/L_{\rm int}^0\sim(L_{\rm dep}\tau_{\rm dep}/L_{\rm eq})^{-(1-\beta)}$, which we rewrite as
\begin{equation}\label{eq:short}
\frac{L_{\rm int}}{L_{\rm int}^0}\sim\left(1+\frac{L_{\rm dep}\tau_{\rm dep}}{L_{\rm eq}}\right)^{-(1-\beta)},
\end{equation}
in order to interpolate with the $L_{\rm dep}\tau_{\rm dep}/L_{\rm eq}\lesssim 1$ regime.

Equation \eqref{eq:short} shows that the figure of merit of the heat source, in terms of its influence on the internal luminosity, is $L_{\rm dep}\tau_{\rm dep}/L_{\rm eq}$, which must satisfy $L_{\rm dep}\tau_{\rm dep}/L_{\rm eq}\gtrsim 1$ for a significant effect. For example, if the heat is deposited at $\tau_{\rm dep}\sim 1$, then $L_{\rm dep}\gtrsim L_{\rm eq}$ is needed. This result is intuitive, since the equilibrium luminosity itself is deposited at $\tau\sim 1$. If, on the other hand, the heat is deposited at the top of the convective region, $\tau_{\rm dep}\sim\tau_{\rm rad}$, then, using Equation \eqref{eq:L_U}, $L_{\rm dep}\gtrsim L_{\rm int}^0$ is sufficient. Although not evident from Equation \eqref{eq:short}, $L_{\rm dep}>L_{\rm int}^0$ is required for a significant effect \citep[see also][]{GuillotShowman2002}, even for deposition in the convective region  $\tau_{\rm dep}>\tau_{\rm rad}$. Otherwise, the deposited power simply replaces part of the internal luminosity as the convectively delivered heat which maintains the radiative profile at the envelope, and the temperature profile remains unchanged (with respect to $L_{\rm dep}=0$). In this case, the luminosity decreases by $L_{\rm dep}$. 

The secondary convective region, described in this section, was also noted in numerical works \citep{GuillotShowman2002,Batygin2011,WuLithwick2013}, for high dissipated power.

\subsection{Cooling and Equilibrium}
\label{sec:cooling_eq}

Substantial additional heating raises the effective equilibrium temperature, and therefore the inner radiative-convective boundary temperature above $T_0$, as seen in Equation \eqref{eq:U_eq_eff_big}. Under the influence of such a heat source, using Equation \eqref{eq:short}, the cooling Equation \eqref{eq:cooling_insolated} in the insolated regime changes to
\begin{equation}\label{eq:cooling_dep}
\begin{split}
T_c(t)&\propto\left(1+\frac{L_{\rm dep}\tau_{\rm dep}}{L_{\rm eq}}\right)^{\beta(1-\beta)/(4-\beta-\beta b)}
\\&=\left(1+\frac{L_{\rm dep}\tau_{\rm dep}}{L_{\rm eq}}\right)^{0.19},
\end{split}
\end{equation}
slowing down the planet's cooling, with the power calculated using the $\rm H^-$ opacity values $\beta=0.35$ and $b=7$, following the discussion in Section \ref{sec:irradiated}. 

Equilibrium is reached when $U_c\approx U_{\rm eq}^{\rm eff}$ (see Section \ref{sec:irradiated}). According to Equations \eqref{eq:tau_rad} and \eqref{eq:U_eq_eff_big}, the required heating intensity for this scenario is
\begin{equation}\label{eq:Ldep_eq}
\frac{L_{\rm dep}\tau_{\rm dep}}{L_{\rm eq}}\sim\left(\frac{U_c}{U_{\rm eq}}\right)^{1/\beta}\sim\frac{\tau_c}{\tau_{\rm rad}},
\end{equation}
with $\tau_{\rm rad}$ denoting, as before, the original radiative-convective transition, without additional heating. We may rewrite Equation \eqref{eq:Ldep_eq}, using Equation \eqref{eq:L_U}, as $L_{\rm dep}\tau_{\rm dep}\sim L_{\rm int}^0\tau_c$, showing that, as intuitively expected, a deposition of order $L_{\rm int}^0$ forces equilibrium, when the deposition is at the center.
For a given heat source, Equation \eqref{eq:Ldep_eq} defines the equilibrium central temperature $T_c=T_{\rm eq}^{\rm eff}\equiv(U_{\rm eq}^{\rm eff}/a)^{1/4}$, for which evolutionary cooling not only slows down, but stops entirely
\begin{equation}\label{eq:Uc_eq}
\frac{T_c}{T_{\rm eq}}=\frac{T_{\rm eq}^{\rm eff}}{T_{\rm eq}}\sim\left(1+\frac{L_{\rm dep}\tau_{\rm dep}}{L_{\rm eq}}\right)^{\beta/4},
\end{equation} 
where we have again interpolated with the weak heating regime. 

If, due to very intensive (or deep) heating, $T_{\rm eq}^{\rm eff}\gtrsim 10^4{\rm K}$, the opacity behaves according to the Kramers law, with $b=-3.5$ and $\beta>1$, prohibiting convection (see Section \ref{sec:irradiated}). In this case, the planet becomes fully radiative from $\tau_{\rm dep}$ inward, even before equilibrium is reached. The internal luminosity in this case is given by Equation \eqref{eq:diff}, $L_{\rm int}\approx(4\pi R^2c/3)(U_c/\tau_c)$. Combining this result with Equation \eqref{eq:cooling}, we find that the cooling time is $t\propto T_c^{-(3-b)}=T_c^{-6.5}$, with hotter (more inflated) planets reaching equilibrium first (planets with radii close to $2R_J$ cool even faster, due to the dependence of the opacity on the density). This regime corresponds to the equilibrium-inflation paradigm of hot Jupiters, which typically involves very deep heating \citep{GuillotShowman2002,Chabrier2004,Burrows2007,Liu2008,SpiegelBurrows2013}.

\section{Effect of Heating on Planet Radius}
\label{sec:radius}

In Sections \ref{sec:irradiated} and \ref{sec:deposition} we found the effects of stellar irradiation and additional heat deposition on the planet luminosity. The diminished luminosity results in slower cooling, and therefore a higher bulk temperature and larger radius at a specific age, in addition to a larger equilibrium radius. For a rough estimate of this inflation, we introduce a simple model which relates the plant's radius $R$ to its central temperature $T_c$.

We start by writing a simple equation of state (EOS)
\begin{equation}\label{eq:eos}
P=K\rho^{5/3}\left[1-\left(\frac{\rho_0}{\rho}\right)^{1/3}\right]+\frac{\rho}{m_p}k_BT,
\end{equation}
where $K\sim h^2/(m_em_p^{5/3})$ is the electron degeneracy term, and $(\rho_0/m_p)^{1/3}\sim m_ee^2/h^2\sim a_0^{-1}$ is the electrostatic correction \citep[$h$ denotes the Planck constant, $m_e$ and $e$ denote the electron mass and charge, respectively, and $a_0$ is the Bohr radius; see, e.g.,][where we assume a mixture of hydrogen and helium, so the atomic weight and charge are approximately 1]{Padmanabhan2001}. The rightmost term is the ideal gas pressure of the non-degenerate ions. We are interested in the regime where the degeneracy parameter $\theta\equiv(\rho k_BT/m_p)/K\rho^{5/3}\sim 10^{-1}$, which is adequate for HD 209458b \citep[see][]{Guillot2005}. In this regime, the electrons are almost completely degenerate, and the ion thermal term is a small correction. Since the degeneracy pressure is inversely related to the particle mass, the ions are non-degenerate. The thermal contribution of the degenerate electrons is quadratic in $\theta$ \citep[see, e.g.,][]{Chandra39}, and is therefore negligible. This EOS is also used by \citet{ArrasBildsten2006}.

We proceed by replacing the profiles $\rho(r),T(r),P(r)$, with the characteristic values $\rho_c,T_c,P_c$. This simplification is justified by noting that the entire planet (except for the radiative isothermal envelope) can be described approximately with a single almost degenerate polytrope. The distinction between a degenerate core and a non-degenerate envelope is insignificant, since the temperature profile is almost parallel to lines of constant $\theta$ \citep{Guillot2005}. A similar argument renders the distinction between a core without electrostatic corrections, and an envelope with significant corrections, also insignificant in comparison with the bulk electrostatic correction. For a given central temperature $T_c$, the contribution of the isothermal ($T\approx T_{\rm eq}$) radiative envelope to the radius is negligible, because it scales with the temperature, and $T_{\rm eq}\ll T_c$. We emphasize that the quasi-degenerate treatment in this section only holds for mild inflations, and should be taken only as an order of magnitude estimate for extremely inflated planets, with radii $\sim 2R_J$. In these cases, numerical radius-temperature relations are more adequate, as detailed below.

In equilibrium, the pressure of Equation \eqref{eq:eos} balances the gravitational pressure, of order $GM^2/R^4$ ($G$ is the gravitational constant),
\begin{equation}\label{eq:prs_grav}
K\rho_c^{5/3}\left[1-\left(\frac{\rho_0}{\rho_c}\right)^{1/3}\right]+\frac{\rho_c}{m_p}k_BT_c\sim GM^{2/3}\rho_c^{4/3}.
\end{equation}
For small variations of the temperature, differentiating Equation \eqref{eq:prs_grav}, while holding the planet mass constant, yields a linear relation
\begin{equation}\label{eq:drho_const_m}
\left.\frac{dR}{R}\right|_M=-\frac{1}{3}\left.\frac{d\rho_c}{\rho_c}\right|_M=\frac{\frac{\rho_c}{m_p}k_BdT_c}{K\rho_c^{5/3}-\frac{\rho_c}{m_p}k_BT_c}.
\end{equation}
In the degenerate limit ($\rho_ck_BT_c/m_p\ll K\rho_c^{5/3}$), the radius increases with temperature, while in the non-degenerate limit ($\rho_ck_BT_c/m_p\gg K\rho_c^{5/3}$), the planet shrinks as the temperature increases. As mentioned above, the non-degenerate limit corresponds to the initial contraction phase, accompanied by an increase in temperature, while the degenerate limit describes the later cooling of the planet, which is accompanied by a decrease in radius. The inflated hot Jupiters are on the late-time cooling branch of this scenario \citep[the transition is at a radius around $4R_J$, see][]{Guillot96}.

We make a linear approximation for low temperatures, using Equations \eqref{eq:prs_grav} and \eqref{eq:drho_const_m}
\begin{equation}\label{eq:delta_r_t}
\begin{split}
\Delta R&\approx\frac{R^2}{GMm_p}k_BT_c\left[1+\left(M_J/M\right)^{2/3}\right]^{-1}
\sim\frac{k_BT_c}{m_pg}
\\&\approx 0.1 R_J\left(\frac{T_c}{10^4{\rm K}}\right),
\end{split}
\end{equation}
with $\Delta R\equiv R-R_0$ denoting the inflation relative to the zero-temperature radius $R_0$, $g$ is the surface gravity, and noting that Jupiter's mass is close the well-known inversion of the zero-temperature mass-radius relation $M_J\sim(K/G)^{3/2}\rho_0^{1/2}$ \citep{Padmanabhan2001}. Since we focus on hot Jupiters with $M\approx M_J$, the electrostatic correction term $M_J/M$ does not influence our order of magnitude estimate. This relation fits well numerical radius-temperature curves \citep[for example,][Table 1]{Burrows97}, up to a radius of $\sim 1.3R_J$, where the linear approximation breaks down (since it assumes $\Delta R\ll R_0\approx R_J$). See \citet{ArrasBildsten2006} for a similar linear relation between the radius change and the central temperature.

Although Equation \eqref{eq:delta_r_t} was derived for a constant planet mass, it is also valid (up to numerical factors) when the planet surface gravity $g$, instead of the planet mass is kept constant. In this case, the gravitational pressure in Equation \eqref{eq:prs_grav} may be written as $g^2/G$, and the radius increase is found by differentiating $dR/R|_g=-d\rho_c/\rho_c|_g$. This scenario is interesting for comparison with the results of \citet{SpiegelBurrows2013}, who studied models with constant surface gravity.

\section{Comparison to Numerical Calculations}
\label{sec:comparison}

\citet{GuillotShowman2002} and \citet{Baraffe2003} find that less heat dissipation is needed to inflate hot Jupiters if the heat is deposited in deeper mass fractions of the planet \citep[see also][]{WuLithwick2013}. In this section we compare our estimates to the systematic work of \citet{SpiegelBurrows2013}, henceforth SB2013, who studied the effects of different amounts of additional heat, deposited at different pressure levels.

In their Section 3.3, specifically Figure 4, SB2013 deposit different fractions of the incident flux $F_{\rm dep}/F_{\rm eq}=L_{\rm dep}/L_{\rm eq}$ at different pressures inside a planet similar to HD 209458b. For each value of additional power and deposition pressure, SB2013 present the inflated planet radius. We emphasize that, in their Section 3.3, SB2013 concentrate on planets which have not yet reached equilibrium, but are still contracting. Moreover, they compare models with the same effective temperature, defined as $T_{\rm eff}\equiv(L_{\rm int}/\sigma 4\pi R^2)^{1/4}$, and the same surface gravity, resulting in models with different ages and masses. 

By combining Equations \eqref{eq:Lint_cases} and \eqref{eq:short}, we relate the central temperature to the deposited power,
\begin{equation}\label{eq:T_c_Ldep}
\frac{T_c}{T_{\rm eq}}\sim\left[\tau_0\left(\frac{T_{\rm eq}}{T_0}\right)^b\frac{L_{\rm int}}{L_{\rm eq}}\left(1+\frac{L_{\rm dep}\tau_{\rm dep}}{L_{\rm eq}}\right)^{1-\beta}\right]^{\beta/(4-\beta b)}.
\end{equation}
The inflated planet radius can be related to the deposited power by combining Equation \eqref{eq:T_c_Ldep} with Equation \eqref{eq:delta_r_t}
\begin{equation}\label{eq:delta_R0}
\begin{split}
\Delta R&=\frac{k_BT_{\rm eq}}{m_pg}\left[\tau_0\left(\frac{T_{\rm eff}}{T_{\rm eq}}\right)^4\left(\frac{T_{\rm eq}}{T_{\rm 0}}\right)^b\right]^{\beta/(4-\beta b)}
\\&\times\left(1+\frac{L_{\rm dep}\tau_{\rm dep}}{L_{\rm eq}}\right)^{\beta(1-\beta)/(4-\beta b)}
\\&\equiv\Delta R_0\left(1+\frac{L_{\rm dep}\tau_{\rm dep}}{L_{\rm eq}}\right)^{\beta(1-\beta)/(4-\beta b)},
\end{split}
\end{equation}
with $\Delta R_0=0.27R_J$ denoting the expansion without extra heating. This numeric value is estimated by substituting the parameters $g=10^3\textrm{ cm s}^{-2}$ and $T_{\rm eff}=180\rm K$, which are kept constant in SB2013, the observed temperature $T_{\rm eq}=1300K\gg T_{\rm eff}$ \citep{Crossfield2012}, and taking the adequate $\beta=0.35$ and $b=7$.
The model of SB2013 for HD209458b is indeed inflated by $\Delta R_0=1.25R_J-0.95R_J=0.30R_J$, due to the irradiation of the parent star, even without extra power sources \citep[see also][]{Baraffe2003,Burrows2007}, in agreement with our estimate. The zero-temperature radius $R_0=0.95R_J$ is estimated using \cite{Guillot2005} and \citet{Baraffe2010}, for the given surface gravity. An alternative estimate of the inflation may be given by relating the central temperature, found in Equation \eqref{eq:T_c_Ldep}, to $\Delta R$, using the numerically calculated $R(T_c)$ curve given in \citet{Burrows97}. This analysis is more accurate at large expansions.

In Figure \ref{fig:compare_log} we compare the results of SB2013 (their Figure 4) to the equation
\begin{equation}\label{eq:compare}
\Delta R_{\rm dep}=\Delta R_0\left[\left(1+\eta\frac{L_{\rm dep}\tau_{\rm dep}}{L_{\rm eq}}\right)^\delta-1\right],
\end{equation}
with $\Delta R_{\rm dep}\equiv\Delta R-\Delta R_0$ denoting the expansion relative to the radius without extra heat deposition ($1.25R_J$ in our case), $\Delta R_0=0.30R_J$, $\eta$ an order unity fitting parameter, and $\delta=\beta(1-\beta)/(4-\beta b)\approx 0.15$ from Equation \eqref{eq:delta_R0}. The fraction of incident flux $L_{\rm dep}/L_{\rm eq}$ is given by SB2013, and the translation of deposition pressure to optical depth is done approximately by $P/g\approx\tau/\kappa$, with $\kappa$ taken from \citet{ArrasBildsten2006}, and uncertainties going into the fitting parameter $\eta$. The parameters in Figure \ref{fig:compare_log}, $\eta=\delta=0.2$, are not, strictly speaking, a best fit, but only an illustration. A second fit, with $\eta=0.25$ and the predicted $\delta=0.15$, is also provided. This fit, however, incorporates the more accurate nonlinear radius correction \citep{Burrows97} to Equation \eqref{eq:compare}.

While our model is not exact, as seen in Figure \ref{fig:compare_log}, it does manage to describe the main trends of atmospheric heating, and, given the general uncertainties and approximations of the model, it fits the numerical calculations reasonably well over several orders of magnitude in pressure level and flux. When the realistic radius-temperature relation is used, the analytically predicted power $\delta=0.15$ fits the results adequately. With the approximated linear radius-temperature relation, however, we fit $\delta=0.2$ to the numerical results. 

\ifpdf

\else

\begin{figure}[tbh]
\epsscale{1} \plotone{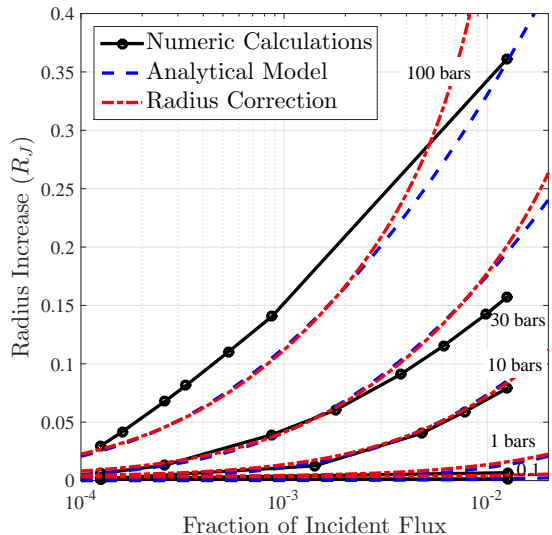}
\caption{Comparison of our prediction for planet inflation (relative to $1.25R_J$), due to heating (expressed as a fraction of the incident flux) at different pressure levels, with \citet{SpiegelBurrows2013}, for constant surface gravity $g=10^3\textrm{ cm s}^{-2}$, effective temperature $T_{\rm eff}=180\rm K$, and incident irradiation of HD 209458b. Our analytical model (dashed blue lines) is given by Equation \eqref{eq:compare}, with fitted parameters $\eta=\delta=0.2$. A second model (dot-dashed red lines), which takes into account the nonlinear radius-temperature relation, is also provided, with a fitted $\eta=0.25$, and the analytically derived $\delta=0.15$. The results of SB2013 (solid black lines, marked with circles) are taken from their Figure 4.  
\label{fig:compare_log}}
\end{figure}

\fi

In the ``weak heating'' regime $L_{\rm dep}\tau_{\rm dep}/L_{\rm eq}<1$, it is possible to linearize Equation \eqref{eq:compare}
\begin{equation}\label{eq:lin_approx}
\Delta R_{\rm dep}\approx\delta\eta\Delta R_0\frac{L_{\rm dep}\tau_{\rm dep}}{L_{\rm eq}}=\zeta\Delta R_0\frac{L_{\rm dep}\tau_{\rm dep}}{L_{\rm eq}},
\end{equation}
which has the advantage of a single fitting parameter $\zeta\equiv\delta\eta\approx 0.04$, allowing for a more intuitive interpretation of the results.

\section{Conclusions}
\label{sec:conclusions}

In this work we presented a simplified analytical model for irradiated giant gas planets, which includes an additional heat source, deposited at an arbitrary depth inside the planet's atmosphere. 

For an irradiated planet with no extra heat sources, we found useful approximate scaling relations for the internal luminosity $L_{\rm int}$ of the planet, and for the radiative-convective boundary, marked by an optical depth $\tau_{\rm rad}$. Explicitly, the radiative-convective boundary and the internal luminosity are given by  
\begin{equation}\label{eq:conc_irradiated}
\frac{L_{\rm eq}}{L_{\rm int}}\sim\tau_{\rm rad}\sim\tau_c\left(\frac{U_{\rm eq}}{U_c}\right)^{1/\beta},
\end{equation}  
with $L_{\rm eq}$, $U_{\rm eq}\propto T_{\rm eq}^4$ denoting the equilibrium luminosity and radiation energy density imposed by the stellar irradiation, respectively ($T_{\rm eq}$ is the equilibrium temperature). $U_c$ is the central radiation energy density of the planet, and $\tau_c$ represents its optical depth (with opacities extrapolated from the outer layers of the planet inward). The power $\beta\approx 0.35$ is related to the adiabatic index of the planet and to the opacity near the planet's edge. In addition, we identified the condition $\tau_{\rm rad}\sim 1$, as the boundary separating insolated (irradiated) planets ($\tau_{\rm rad}>1$) from isolated ones ($\tau_{\rm rad}<1$). These relations serve as an intuitive interpretation of well-known results \citep{Guillot96,Burrows2000,ArrasBildsten2006,FortneyNettelmann2010,YoudinMitchell2010,SpiegelBurrows2013}.

We addressed the effects of possible additional heat sources, parametrized by additional luminosity $L_{\rm dep}$, deposited at an optical depth $\tau_{\rm dep}$. Such heat sources lower the internal luminosity of the planet, slow its evolutionary cooling, and are therefore candidates for solving the puzzle of over-inflated hot Jupiters \citep[see][and references within]{Baraffe2010,FortneyNettelmann2010,SpiegelBurrows2013,Baraffe2014}. According to our model, the figure of merit in this context is $L_{\rm dep}\tau_{\rm dep}/L_{\rm eq}$, with $L_{\rm dep}\tau_{\rm dep}/L_{\rm eq}\gtrsim 1$ necessary for a significant effect. Concretely, an outer convective layer is formed in this case, between optical depths $L_{\rm eq}/L_{\rm dep}$ and $\tau_{\rm dep}$, which is equivalent to an enhanced stellar irradiation
\begin{equation}\label{eq:conc_Leq_eff}
\frac{L_{\rm eq}^{\rm eff}}{L_{\rm eq}}=\left(\frac{T_{\rm eq}^{\rm eff}}{T_{\rm eq}}\right)^4\sim\left(1+\frac{L_{\rm dep}\tau_{\rm dep}}{L_{\rm eq}}\right)^\beta.
\end{equation}
The internal luminosity is therefore reduced from a value of $L_{\rm int}^0$ without extra heating to, roughly,
\begin{equation}\label{eq:conc_luminosity}
\frac{L_{\rm int}}{L_{\rm int}^0}\sim\left(1+\frac{L_{\rm dep}\tau_{\rm dep}}{L_{\rm eq}}\right)^{-(1-\beta)},
\end{equation} 
with a further threshold of $L_{\rm int}^0$ required for a significant effect.

Since the inflation of a planet, relative to its zero-temperature radius, is proportional to its central temperature, these relations provide an intuitive explanation to the dependence of inflation on the deposition depth \citep{GuillotShowman2002,Baraffe2003,WuLithwick2013}, and to the large equilibrium radii, in the case of very deep heating \citep{GuillotShowman2002,Burrows2007,Liu2008,SpiegelBurrows2013}.

Quantitatively, we compared our model to the recent parameter survey of \citet{SpiegelBurrows2013}. The model, though not exact, fits the numerical results adequately, considering its simplifying assumptions and approximations.

The scaling laws presented in this work, combined with potential observational correlations \citep[see, e.g.,][]{Laughlin2011, Schneider2011}, may be used to rule out or to favor specific energy sources, or even the entire additional heat source paradigm.

\acknowledgements This research was partially supported by ISF, ISA and iCore grants, and a
Packard Fellowship.
RS would like to thank the Max Planck Institute for Extraterrestrial Physics and the Humbolt Foundation for support and worm hospitality while this research was completed. We thank
Peter Goldreich for insightful discussions, and the anonymous referee for helpful comments.

\bibliographystyle{apj}

\end{document}